\documentclass[11pt,a4paper]{article}

\usepackage{arxiv} 

\usepackage[utf8]{inputenc} 
\usepackage[T1]{fontenc}    
\usepackage{authblk}        
\usepackage{url}            
\usepackage{booktabs}       
\usepackage{amsfonts}       
\usepackage{nicefrac}       
\usepackage{microtype}      
\usepackage{caption}
\usepackage{subcaption}
\usepackage{comment}
\usepackage{lipsum}
\usepackage[dvipsnames]{xcolor}
\usepackage{multirow}
\usepackage{soul}

\usepackage{graphicx}
\usepackage{afterpage}
\usepackage{array}
\usepackage{booktabs}
\usepackage{amssymb}
\usepackage{amsbsy}
\usepackage{newtxmath}

\newcommand{\ie}[0]{\textit{i.e.}}

\title{Learning Curves for Drug Response Prediction in Cancer Cell Lines}

\author[1,2]{Alexander Partin} 
\author[2,3]{Thomas Brettin}
\author[4]{Yvonne A. Evrard}
\author[1,2]{Yitan Zhu}
\author[1,2]{Hyunseung Yoo}
\author[1,2]{Fangfang Xia}
\author[7]{Songhao Jiang}
\author[1,7]{Austin Clyde}
\author[1,2]{Maulik Shukla}
\author[5]{Michael Fonstein}
\author[6]{James H. Doroshow}
\author[3,7]{Rick Stevens}

\affil[1]{Division of Data Science and Learning, Argonne National Laboratory, Argonne, IL, USA}
\affil[2]{University of Chicago Consortium for Advanced Science and Engineering, University of Chicago, Chicago, IL, USA}
\affil[3]{Computing, Environment and Life Sciences, Argonne National Laboratory, Lemont, IL, USA}
\affil[4]{Frederick National Laboratory for Cancer Research, Leidos Biomedical Research, Inc. Frederick, MD, USA}
\affil[5]{Biosciences Division, Argonne National Laboratory, Lemont, IL, USA}
\affil[6]{Division of Cancer Therapeutics and Diagnosis, National Cancer Institute, Bethesda, MD, USA}
\affil[7]{Department of Computer Science, The University of Chicago, Chicago, IL, USA}

\begin{document}
\maketitle

\begin{abstract}

Motivated by the size and availability of cell line drug sensitivity data, researchers have been developing machine learning (ML) models for predicting drug response to advance cancer treatment. As drug sensitivity studies continue generating drug response data, a common question is whether the proposed predictors can further improve the generalization performance with more training data.
We utilize empirical learning curves for evaluating and comparing the data scaling properties of two neural networks (NNs) and two gradient boosting decision tree (GBDT) models trained on four cell line drug screening datasets. The learning curves are accurately fitted to a power law model, providing a framework for assessing the data scaling behavior of these predictors.
The curves demonstrate that no single model dominates in terms of prediction performance across all datasets and training sizes, thus suggesting that the actual shape of these curves depends on the unique pair of an ML model and a dataset. The multi-input NN (mNN), in which gene expressions of cancer cells and molecular drug descriptors are input into separate subnetworks, outperforms a single-input NN (sNN), where the cell and drug features are concatenated for the input layer. In contrast, a GBDT with hyperparameter tuning exhibits superior performance as compared with both NNs at the lower range of training set sizes for two of the tested datasets, whereas the mNN consistently performs better at the higher range of training sizes. Moreover, the trajectory of the curves suggests that increasing the sample size is expected to further improve prediction scores of both NNs. These observations demonstrate the benefit of using learning curves to evaluate predictors, providing a broader perspective on the overall data scaling characteristics.
A fitted power law learning curve provides a forward-looking metric for analyzing prediction performance and can serve as a co-design tool to guide experimental biologists and computational scientists in the design of future experiments in prospective research studies.

\end{abstract}


\section{Background} \label{sec:intro}

Human cancer cell lines remain a primary cancer-mimicking environment in a laboratory setting for understanding the molecular biology of this complex disease \cite{Sharma2010, Gillet2013, Ben-David2018}. In the search for anticancer treatments, in vitro drug sensitivity assays serve as a standard, high-throughput experimental platform for measuring the response of cancer cells to drug treatments. The standardized protocols of sensitivity assays, along with rapid improvement of technologies for genomic profiling, have led researchers to generate large pharmacogenomic drug response datasets for anticancer drug discovery \cite{yang2013gdsc, ludlow2015ctrp, grever1992nci60}. Considering the scale and diversity of tumors and compounds in these datasets, machine learning (ML) techniques have become a natural fit for analytically predicting the response of cell lines to drug treatments. By maneuvering through a landscape of computational approaches and numerical representations of tumors and drugs, researchers strive to develop highly predictive ML drug response models \cite{costello2014, DeNiz2016, adam2020}. Demonstrating the accuracy and robustness of predictors is essential in order to identify their potential utility for clinical applications in cancer treatment including precision oncology and drug repurposing.

In ML-driven cancer research, a common question is whether existing predictive models can be further improved with more training data. Given recent advances in artificial neural networks (NNs), deep learning (DL) methods have become a favorite approach across a variety of scientific disciplines for discovering hidden patterns in large volumes of complex data. This trend is also observed in medical applications, including the prediction of drug response in cancer cell lines \cite{Rampasek2019a, Manica2019b, Bazgir2020, Zhu2020EnsembleTL, Vougas2019}. Regardless of the learning algorithm, supervised learning models are expected to improve generalization performance with increasing amounts of high-quality labeled data. Generalization performance refers to the aggregated accuracy of model predictions on a set of unseen data samples. Analytically estimating the learning capacity of models is a challenging task. Alternatively, given a dataset and a learning algorithm, the projected improvement of predictions with increasing number of training samples can be empirically estimated by using \textbf{learning curves}.

A learning curve is a plot of the generalization performance of a predictor as a function of training set size (Fig. \ref{fig:lrn_crv_intro}). These curves have been explored as an efficient method for modeling the power law relationship, $s(m)\propto am^{b}$, between the generalization score $s$ (such as generalization error or accuracy) and the number of training samples $m$, where $a$ and $b$ are two parameters of the power law model. The power law characteristics of learning curves can provide insights into the data scaling behavior of drug response predictors, which otherwise could not be investigated by merely analyzing single-value performance measures obtained with the full training set size.

A main bottleneck of utilizing learning curves, however, is often the limited availability of sufficient computational resources for performing the analysis. Particularly challenging is analysis with DL models and large datasets because of the large computational cost. While learning curves have been explored in a variety of small-scale applications with classical ML \cite{Cortes1993, Mukherjee2003, Last2007, Figueroa2012}, only a few recent studies have applied DL methods to large benchmark datasets in vision and text applications \cite{Sun2017, Hestness2017, Rosenfeld2020}. To the best of our knowledge, learning curves of drug response predictors have not been previously explored.

In this paper, we develop a practical methodology for generating learning curves of predictive models using both NNs and classical ML algorithms. Specifically, we utilize this method for evaluating drug response predictors implemented with gradient boosting decision tree (GBDT) models and NNs, where each model is trained on four large drug response datasets of cancer cell lines. Using learning curves, we investigate the data scaling trajectory of predictors for each pair of a dataset and learning algorithm. We also are interested in comparing the performance of a powerful classical ML algorithm (GBDT) and DL models across a range of training sizes. To accomplish our objectives, we develop an efficient computational workflow, leveraging high-performance computing (HPC) resources to conduct the large-scale analysis.

\section{Learning curves} \label{sec:lrn_crv}

Theoretical \cite{Amari1992, Haussler1996} and empirical \cite{Cortes1993, Hestness2017, Rosenfeld2020} studies demonstrate that learning curves of predictive models are characterized by a power law relationship between the training set size $m$ and the generalization score $s$,
\begin{equation} \label{eq:power_law}
s(m) = am^{b}+c,
\end{equation}
where $\beta=(a,b,c)$ is the set of the power law parameters. The parameters in $\beta$ determine the shape of the curve and typically vary for individual combinations of learning algorithm, prediction task, and data.

\begin{figure*}[h]
  \centering
  \begin{subfigure}[]{0.47\textwidth}
    \centering
    \includegraphics[width=\textwidth]{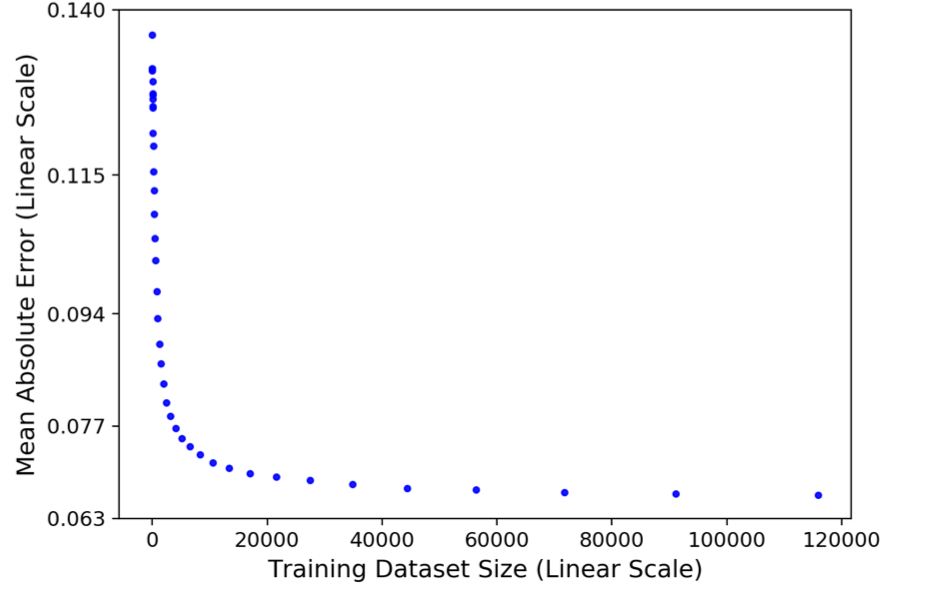}
    \caption{}
    \label{fig:lc_lin_scale}
  \end{subfigure}
  \begin{subfigure}[]{0.47\textwidth}
    \centering
    \includegraphics[width=\textwidth]{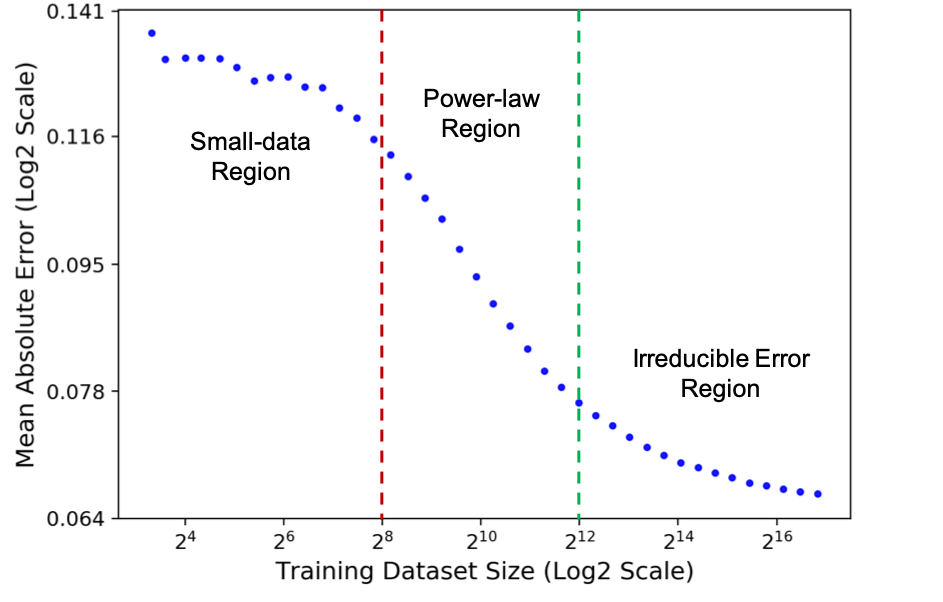}
    \caption{}
    \label{fig:lc_log_scale}
  \end{subfigure}
  \caption{Learning curve plotted on a linear scale in (a) and on a log scale in (b).
  The vertical axis is the generalization score in terms of the mean absolute error of model predictions. Each data point is the averaged prediction error, computed on a test set, of a gradient boosting decision tree (GBDT) that was trained on a subset of training samples of the GDSC1 dataset.}
  \label{fig:lrn_crv_intro}
\end{figure*}

An empirical learning curve often exhibits three primary learning regions: small-data, power law, and irreducible error \cite{Hestness2017}. Figure \ref{fig:lrn_crv_intro} provides a glimpse into the experimental results discussed later. All three learning regions of the power law expression in Eq. (\ref{eq:power_law}) are distinguishable on the \textit{log-log} plot in Fig. \ref{fig:lc_log_scale}. Alternatively, the linear scale representation in Fig. \ref{fig:lc_lin_scale} visually obscures these learning trends.

The small-data region is attributed to the lack of an appropriate amount of training samples for learning a sufficiently generalizable predictor. This region produces models that perform as good as, or slightly better than, random (or best) guessing and therefore is commonly known as the region of random guess \cite{Hestness2017, Rosenfeld2020}. On a $log$-$log$ plot, the random guessing is characterized primarily by a horizontal flat region, followed by a transition to the power law region.

From random guessing, the curve transitions to the power law region described by the $am^{b}$ term. In this region, the curve maintains a steady trajectory, as shown by the approximately constant slope on the $log$-$log$ plot in Fig. \ref{fig:lc_log_scale}. The parameter $b$, ranging between $0<|b|<1$, is the scaling exponent and determines the steepness of the curve in the power law region \cite{Mukherjee2003, Hestness2017, Anzanello2011}.

As the training size increases, the model starts to exhaust its learning capacity, gradually approaching a plateau, known as the irreducible asymptotic error \cite{Mukherjee2003, Hestness2017, Rosenfeld2020}. The constant term $c$ in Eq. (\ref{eq:power_law}) accounts for a smooth transition from the power law region into this plateau. When this convergence region becomes apparent in the plot, it implies that the model is not expected to significantly improve with more training data, providing researchers with valuable information for future directions in their attempts to improve model predictions.

Learning curves provide intuitive insight into the data scaling behavior of predictors, as opposed to single-value performance measures obtained with the entire set of training samples. The shape of these curves facilitates comparison between ML models by illustrating a global trajectory of model improvement. Thus, learning curves can be utilized for quantifying the learning capacity of predictors with increasing amounts of training data. 

\section{Methods} \label{sec:methods}

This section describes the drug response datasets, learning algorithms, training procedures, and methodology for generating and fitting learning curves.

\subsection{Drug response datasets} \label{sec:datasets}
The four datasets used for the experiments are listed in Table \ref{tab:datasets}.
The data comes from public repositories of drug sensitivity studies: the Genomics of Drug Sensitivity in Cancer project, which includes GDSC1 and GDSC2 datasets \cite{yang2013gdsc}; the Cancer Therapeutics Response Portal (CTRP v2) \cite{ludlow2015ctrp}; and the NCI-60 Human Tumor Cell Lines Screen (NCI-60) \cite{grever1992nci60}.

\begin{table}[h]
\caption{Datasets used for training ML models and generating learning curves.}
   \centering
   \begin{tabular}{l | rrrrr}
       \toprule
       Dataset & Responses & Cell Lines & Drugs \\
       \midrule
       GDSC1 & 144,832\hspace{5pt} & 634\hspace{10pt} & 311\hspace{2pt}\\
       GDSC2 & 98,032\hspace{5pt} & 554\hspace{10pt} & 174\hspace{2pt}\\
       CTRP & 254,566\hspace{5pt} & 812\hspace{10pt} & 495\hspace{2pt}\\
       NCI-60 & 750,000\hspace{5pt} & 59\hspace{10pt} & 47,541\hspace{2pt}\\
       \bottomrule
   \end{tabular}
  \label{tab:datasets}
\end{table}

\begin{figure}[h]
  \centering
  \includegraphics[width=0.9\linewidth]{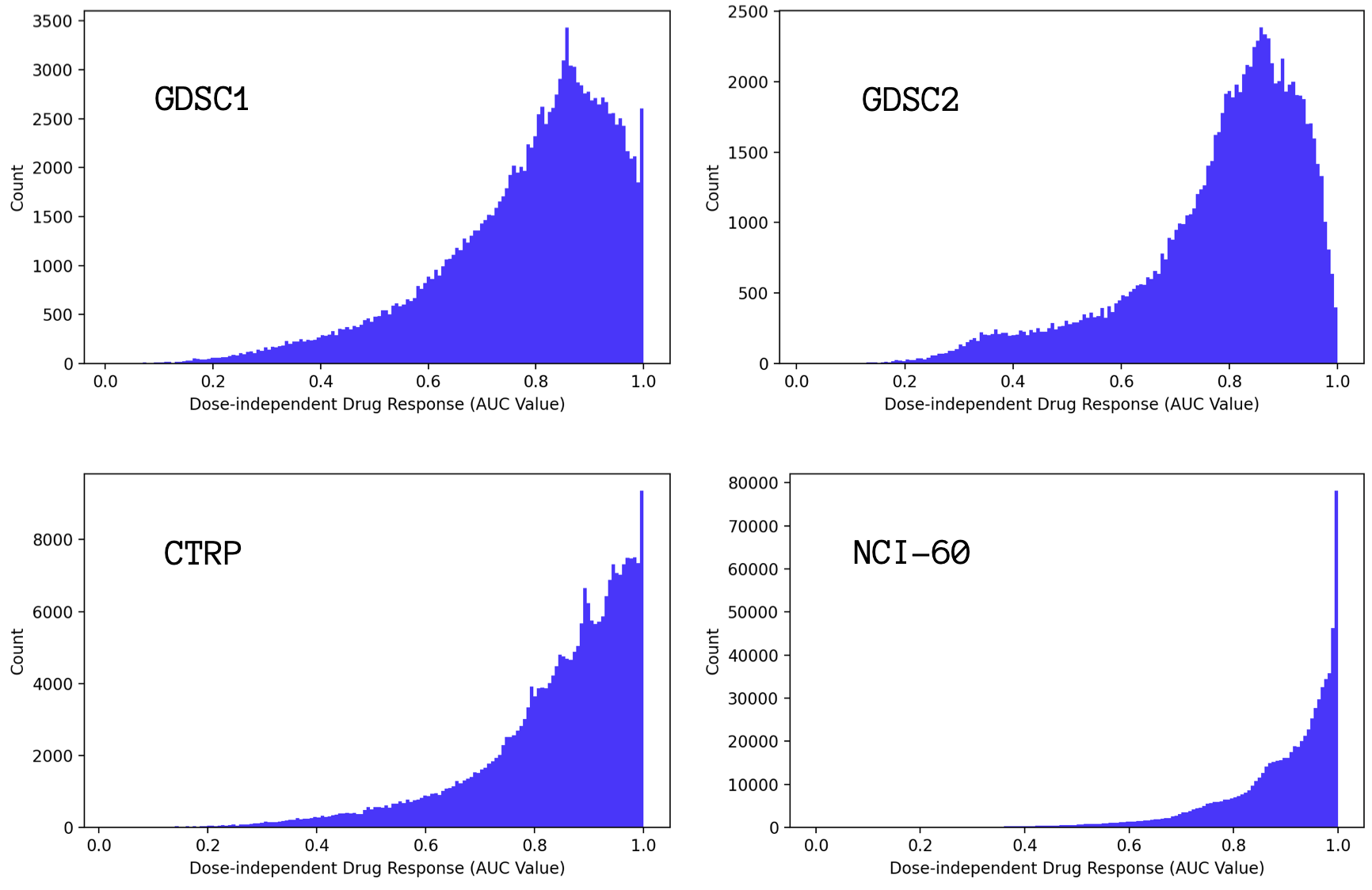}
  \caption{Histograms of dose-independent drug response (AUC) values of the four datasets listed in Table \ref{tab:datasets}.}
  \label{fig:histograms}
\end{figure}

In drug sensitivity data, the drug response of a cancer cell line to a drug treatment is measured by the percentage of viable cells at multiple drug doses. A three-parameter Hill--Slope model was used to fit the dose response curve for each cell-drug pair. To retain high-quality response data, we removed samples in which the ${R^2}$ of the dose response curve fit was lower than 0.3. The fitted curve was used to calculate the area under the curve (AUC) over a dose range of $[10^{-10}$M, $10^{-4}$M$]$. The AUC value was then normalized by the dose range to take a value between 0 and 1. We note that lower AUC values indicate higher growth inhibition (\ie, stronger response to drug treatment). Figure \ref{fig:histograms} shows the histograms of the response values.

The NCI-60 dataset originally included more than 3 million samples, where a sample refers to a drug response of a cell-drug pair. We randomly selected 750,000 samples from the full NCI-60 collection for our analysis. For the GDSC1, GDSC2, and CTRP datasets, we collected all the available samples for which we were able to retrieve or calculate feature representations of cells and drugs.

\subsubsection{Representation of cell lines} \label{sec:cell_fea}

A variety of genomic representations of cancer cells have been used as input features for drug response prediction \cite{xia2018, moli2019, zhu2020coxen}. A number of studies have shown, however, that gene expression exhibits a superior predictive power for modeling drug response in cancer cells \cite{costello2014, xia2018, Jang2014}. For our analysis, we used gene expression data generated by the RNA-Seq technology. The expression data was collected from two public data repositories: NCI-60 and Cancer Cell Line Encyclopedia (CCLE). For the cells in the NCI-60 dataset in Table \ref{tab:datasets}, the RNA-Seq was retrieved from the NCI-60 repository. For the GDSC1, GDSC2, and CTRP datasets, the RNA-Seq was retrieved from the NCI-60 and CCLE repositories by matching the cell line names across the databases.
Drug response samples for which the cell gene expressions were not available in CCLE or NCI-60 were excluded from the datasets.

The retrieved RNA-Seq data, provided in FPKM (fragments per kilobase million) values, was transformed into TPM (transcripts per kilobase million) values. Instead of using expressions of more than 20,000 available genes for modeling drug response, we used the expressions of 976 landmark genes as identified by the Library of Integrated Network-Based Cellular Signatures (LINCS) project \cite{Subra2017lincs}. The LINCS gene set has been shown to comprehensively characterize transcriptomic changes under various chemical and genetic perturbations \cite{Subra2017lincs}.

\subsubsection{Representation of drugs} \label{sec:drug_fea}



Classical ML algorithms such as GBDT ignore the arrangement of features in datasets while utilizing the feature values only. Since we want to compare the learning curves of classical ML and DL models, we used molecular descriptors as drug representations in which the ordering of features is not intended to carry meaningful information. The descriptors were generated by using the Mordred software package \cite{mordred2018}. The full descriptor set comprises 1,826 features, including both 2-D and 3-D molecular structure descriptors. Since most of the 3-D descriptors resulted in invalid (NaN) values for the majority of compounds, we retained only the 2-D descriptors, providing a total of 1,613 drug features.

\subsection{Machine learning for drug response prediction} \label{sec:models}
This section first presents the formulation of drug response prediction as a supervised learning problem and then describes the learning algorithms and model training procedures.

\subsubsection{Drug response prediction as a supervised learning task} \label{sec:math_ml_formulation}

Consider the following definitions for a supervised learning problem of drug response prediction. Given a training set $T$ with $M$ samples, $T=\{\boldsymbol{x}_{i},y_i\}_{i=1}^M$, $\boldsymbol{x}_i$ is a feature vector for cell-drug pair $i$, and $y_i$ is the corresponding dose-independent drug response. The prediction task is to learn a mapping function $f:\mathbb{R}^{C+D}\rightarrow \mathbb{R}$, where $C$ and $D$ are the number of gene expressions and drug descriptors, respectively. Drug response datasets comprise a unique set of cell-drug pairs, $X=[X_{c}\; X_{d}]=\{[\boldsymbol{x}_{c,i}\; \boldsymbol{x}_{d,i}]\}_{i=1}^M$, where $\boldsymbol{x}_{c,i}$ and $\boldsymbol{x}_{d,i}$ are, respectively, the cell and drug feature vectors of the $i^\text{th}$ pair. The algorithm learns the mapping function (\ie, prediction model) by minimizing a regression loss function, which in our analysis is the mean squared error of prediction outcomes.

Note that each cancer cell line was screened against multiple drugs and, vice versa, each drug was tested on multiple cell lines. Thus, although each cell-drug combination is unique in the training set $T$, the feature vectors of individual cells, $\boldsymbol{x}_c$, and drugs, $\boldsymbol{x}_d$, appear multiple times in $T$. The analysis of how this redundancy in feature space affects prediction models and learning curves is beyond the scope of this paper and provides a topic for further investigation.

\subsubsection{Machine learning models and training procedures} \label{sec:models_and_training}

NNs can be designed to enhance learning from a particular feature type of cell or drug \cite{Manica2019b, Ciriano2019Kekule}. In such models, the prediction performance depends on the availability, quality, and diversity of that specific feature type in a training set. In contrast, our primary objective is to gain insight into the overall prediction improvement with an increasing number of training samples. Thus, we refrained from using architectures that focus on learning from specific feature types of cells or drugs.

\begin{figure}[h]
  \centering
  \includegraphics[width=10cm]{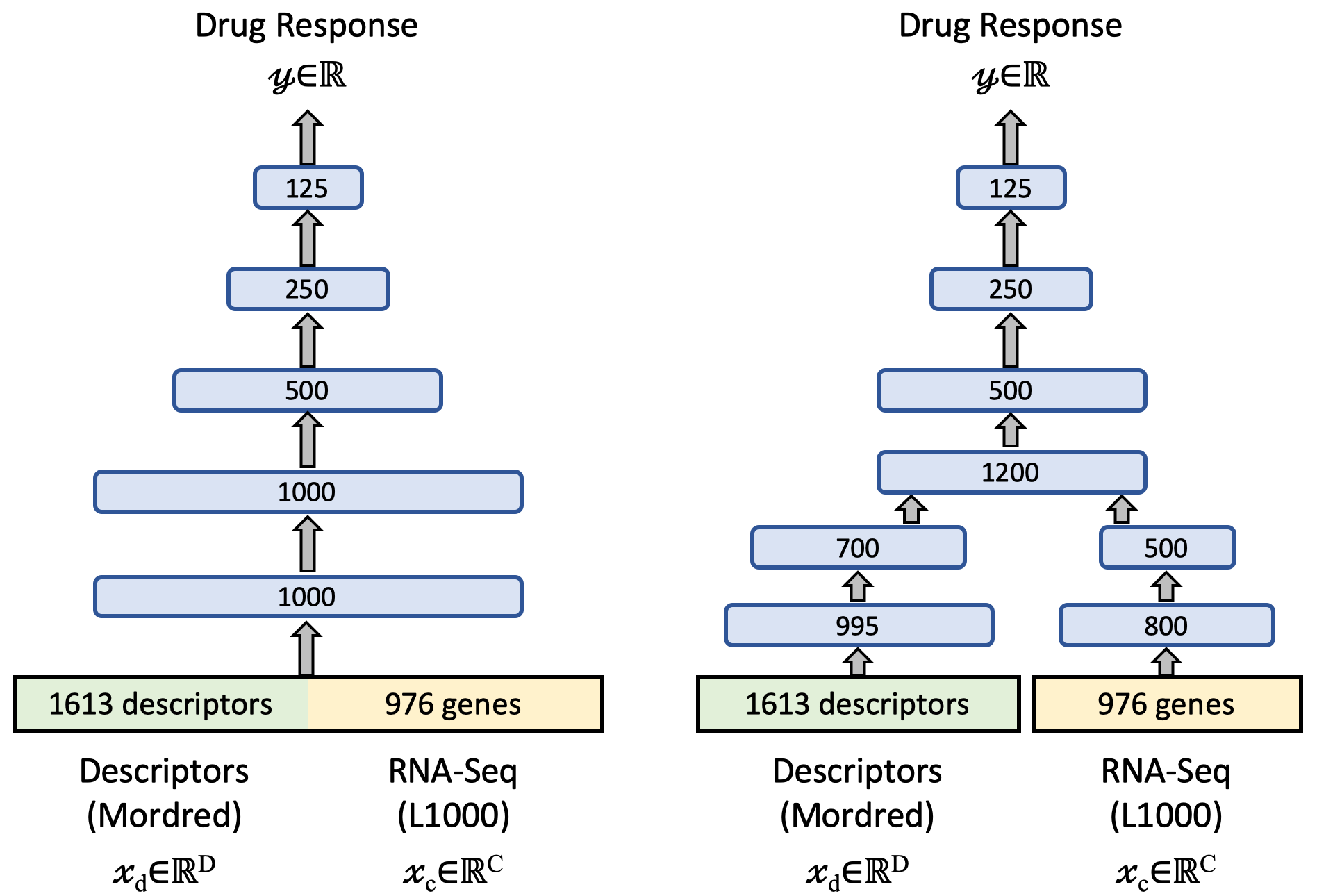}
  \caption{Two neural network architectures used in the analysis: (a) single-input network (sNN, 4.254 million trainable parameters) and (b) multi-input network (mNN, 4.250 million trainable parameters).}
  \label{fig:nn_arch}
\end{figure}

Two NN architectures and one classical ML algorithm were used for the analysis. The two NNs differ primarily in the way the features were fed to the network input, as shown in Fig. \ref{fig:nn_arch}. In the single-input NN (sNN), gene expressions and drug descriptors were concatenated to form an input feature vector. In the multi-input NN (mNN), expression features and descriptors were first encoded by separate subnetworks before being concatenated and subsequently propagated to the output. Both models contain approximately the same number of trainable parameters (sNN: 4.254 million parameters, mNN: 4.250 million parameters). All fully connected layers, excluding the output layer, were followed by batch normalization \cite{batchnorm2015}, ReLu activation, and a dropout layer \cite{dropout2014}. A batch size of 32 and the Adam optimizer \cite{adam2015} were used for model training.

For classical ML, we used the GBDT algorithm, implemented in the LightGBM library \cite{Ke2017LightGBM}. We used two versions of this algorithm: (1) dGBDT, a GBDT with default hyperparameters (HPs), and (2) hGBDT, where we optimized the HPs via a randomized search \cite{Bergstra2012}. GBDT is an ensemble of decision trees in which a series of tree learners is optimized via a gradient descent optimization. Every subsequent tree improves inaccurate predictions of previous learners, boosting the predictive performance of the final model.

The GBDT and NNs were trained with, respectively, the LightGBM \cite{Ke2017LightGBM} and Keras \cite{chollet2015keras} software libraries. The best set of HPs for each combination of a model and dataset was determined based on a randomized HP search by training on 80\% of the available data and validating on the remaining 20\%. Note that default HPs, as provided by the LightGBM library, were used for training dGBDT.

To mitigate overfitting, we used the \textit{early stopping} functionality, available in both LightGBM and Keras. With early stopping, the model training procedure is terminated if the prediction performance on a validation set has not improved for a specified number of training iterations. We set the early stopping parameter to 25 epochs for the NNs and 50 boosting rounds for GBDT models. To guarantee convergence of NNs, we used a sufficiently large number of 500 epochs to ensure that the early stopping function was triggered.

\subsection{Workflow for generating and fitting learning curves} \label{sec:workflow}
\begin{figure}[h]
  \centering
  \includegraphics[width=8.5cm]{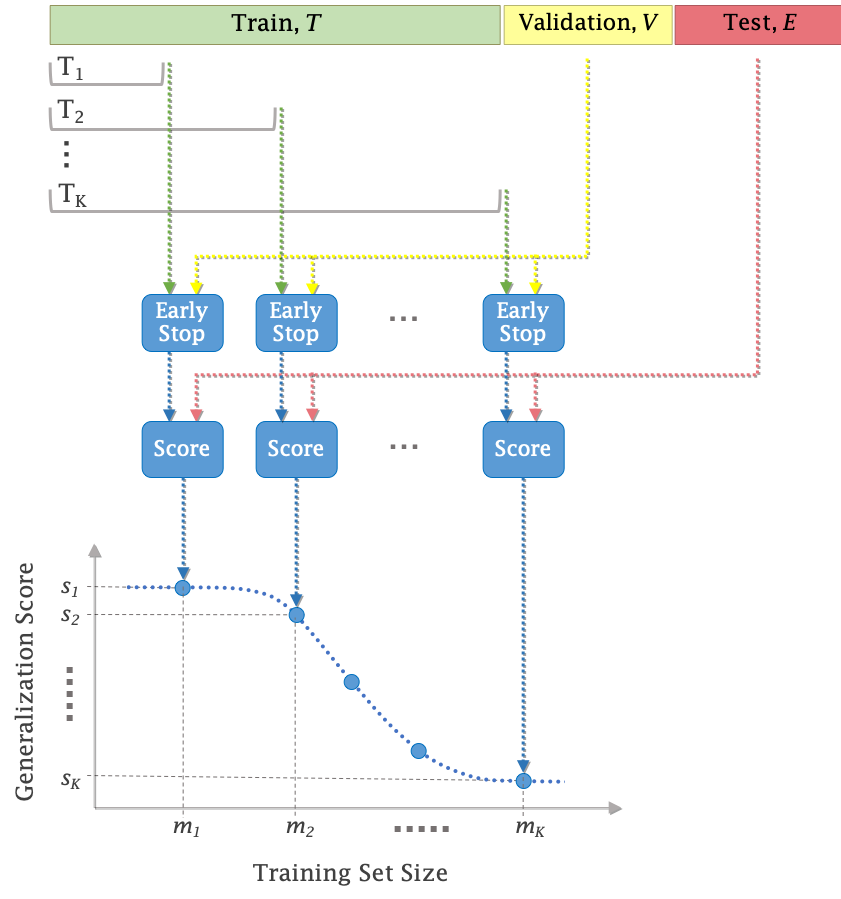}
  \caption{
  Workflow for generating learning curve data, $LC$, for a single split of a dataset. A single dataset split includes three sample sets: training $T$, validation $V$, and test $E$. 
  }
  \label{fig:workflow}
\end{figure}

This section lays out the methodology for generating learning curve data and fitting a power law model. A schematic of the workflow is illustrated in Fig. \ref{fig:workflow}.

\subsubsection{Data partitioning} \label{sec:data_partitioning}

A dataset $D$ is randomly shuffled and split into three disjoint sets: training $T$, validation $V$, and test $E$. Shuffling $D$ before generating the splits increases the likelihood that the three partitions exhibit a similar distribution of drug responses. A total of $N=20$ combinations, $\{T, V, E\}_{n=1}^{N}$, were generated by shuffling $D$ with different random seeds. Each set $\{T, V, E\}$ maintains the same size proportion of (0.8, 0.1, 0.1) as a fraction of the total number of samples $|D|$.


For each of the $N$ splits, we form a set of $K$ training subsets, $\{T_k\}_{k=1}^{K}$, of increasing sizes $m_{k-1}<m_k$, where $m_k$ is the sample size of $T_{k}$. The samples are sequentially pulled from $T$ to form each subset $T_k$, as illustrated in Fig. \ref{fig:workflow}. This process ensures that each subset contains all the training samples from the immediately preceding subset such that $T_{k-1}\subseteq T_k$. The inclusion of the entire subset $T_{k-1}$ in $T_{k}$ mimics the temporal evolution of a research study where new samples are added to an already available dataset. Note that shuffling $D$ at the first step of this data partitioning workflow is essential to enable the sequential sampling from $T$ while eliminating potential biases associated with the original ordering of samples in $D$.

\subsubsection{Generating learning curve data} \label{sec:gen_lc_data}

Once the $N$ data splits and the $K$ training subsets for each split are generated, a total of $N \times K$ prediction models are trained. In order to support early stopping, the training is terminated if the generalization performance on the validation set $V$ has not improved over a predefined number of training iterations. Upon training completion, the model predicts the response for each cell-drug pair in the test set $E$. Note that within a given data split $n$, all $K$ models use the same validation and test sets ($V$ and $E$) irrespective of $m_k$.

The predictions on test set $E$ are aggregated into a single score, $s_k^n$, by computing the mean absolute error of predictions, where ${s_k^n}$ is the generalization error for a subset of size $m_k$ of split $n$. As a result, the \textit{raw} learning curve data, ${LC_{raw}=\{s_k^n\}_{k,n}, k=1,...,K; n=1,...,N}$, is produced. The total number of models trained for each combination of a dataset and a model depends on the values of $N$ and $K$. For most dataset-model pairs, we trained $K=10$ models per split, resulting in a total of 200 models (200 raw $LC$ error scores).

There is no sequential dependency in terms of model training with different subsets and data splits for generating the raw $LC$ data. Thus, the proposed workflow enables the trainings to be distributed across multiple processors. The workflow was parallelized on appropriate platforms depending on the ML software framework. The NNs were trained on the Summit HPC, while the GBDT models were trained a CPU cluster.


\subsubsection{Curve fitting} \label{sec:crv_fit_method}

The raw learning curve data, ${LC_{raw}}$, was aggregated per size $m_k$ by computing the median across the $N$ scores. This averaging produces ${LC_f=\{s_k\}_{k}, k=1,...,K}$, a set of values that was considered next for fitting the power law expression, $s(m) = am^{b}+c$, where $\beta=(a,b,c)$ are the parameters to be estimated. Note that the power law expression in Eq. (\ref{eq:power_law}) primarily accounts for points that span the power law ($am^{b}$ term) and plateau (constant $c$) regions, while mostly excluding the region of small-data. Thus, an inadequate choice of $LC_f$ can lead to a bad fit.

As a first step in producing a reliable fit, we visually identified points in the small-data region and exclude them from $LC_f$ before fitting. To further reduce the effect of small training subsets that are in close proximity with the random guessing, we prioritized the contribution of larger subsets by weighting the least-squares fit (\ie, assigning  larger weights, $\alpha_k$, to larger values of $m_k$).
Specifically, each point in $LC_f$ was assigned a weight that is normalized to the sample size, $\alpha_k=m_k/m_K$ \cite{Figueroa2012, Johnson2018}. The parameter estimates $\hat{\beta}$ were obtained by fitting the remaining $LC_f$ points with a weighted nonlinear least-squares minimization method as described in \cite{Johnson2018}. 

We used the $MAE_{fit}$ and $R_{fit}^2$ as the goodness-of-fit measures. The $MAE_{fit}$ is the mean of absolute value of residuals between the observed scores, $s$, and the estimated scores, $\hat{s}$, where smaller values of $MAE_{fit}$ indicate a closer fit. These measures have been shown as appropriate metrics for evaluating the quality of the power law fit \cite{Last2007, Figueroa2012}.

\section{Results} \label{sec:results}

A single experiment refers to the workflow of generating $LC_{raw}$ and fitting $LC_{f}$ to the power law expression in Eq. (\ref{eq:power_law}) for a pair of a dataset and an ML model. Pairing all the possible combinations results in a total of sixteen experiments. The prediction error scores of models trained with the full training set, $s_K=s(m=m_K)$, are listed in Table \ref{tab:compare_scores}. The dGBDT serves as the baseline model in this comparison and is used to calculate the reduction in error score of other models with
\begin{equation} \label{eq:improvement}
\Delta_s(\alpha) = 100\cdot\frac{s_K(\alpha) - s_K(dGBDT)}{s_K(dGBDT)},
\end{equation}
where $\alpha$ is the evaluated ML model. All models yield lower $s_K$ as compared with dGBDT, exhibiting an improvement of 8\% up to 42\% across the datasets. While these results were expected, the observed improvements render dGBDT as an inadequate baseline for drug response prediction.

\begin{table}[h]
\caption{Prediction scores of all dataset-model combinations.
$s_K$: prediction scores of models trained with the full training set size.
$\Delta_s$: the improvement in prediction error as compared with the baseline dGBDT.
$s(m=2|T|)$: the expected prediction score if the training size is doubled (in parentheses is the percentage reduction in the error score as compared with $s_K$).
$m(s=0.9s_K)$: the training size required to reduce the error score by 10\% (in parentheses is the required increase in sample size as a factor of $|T|$ to achieve the score).
}
   \centering
   \begin{tabular}{ll | ccc | cc}
       \toprule
       Dataset & ML Model & $m_K=|T|$ & $s_K$ & $\Delta_s$ & $s(m=2|T|)$ & $m(s=0.9s_K)$ \\
       \midrule
       \multirow{4}{*}{GDSC1} & dGBDT\hspace{5pt} & \multirow{4}{*}{115,863} & 0.066553 \hspace{2pt} & N/A & 0.066010 (0.82\%) & N/A \\
        & hGBDT\hspace{5pt} && 0.061068 \hspace{2pt} & 8.24\%  & 0.058532 (4.15\%) & 648,331 (x5.6) \\  
        & sNN\hspace{5pt}   && 0.059532 \hspace{2pt} & 10.55\% & 0.056013 (5.91\%) & 347,175 (x3.0) \\
        & mNN\hspace{5pt}   && 0.056904 \hspace{2pt} & 14.50\% & 0.052138 (8.37\%) & 262,810 (x2.7) \\
       \midrule
       \multirow{4}{*}{GDSC2} & dGBDT\hspace{5pt} & \multirow{4}{*}{78,423} & 0.058665 \hspace{2pt} & N/A & 0.058038 (1.07\%) & N/A \\
        & hGBDT\hspace{5pt} && 0.051803 \hspace{2pt} & 11.70\% & 0.049562 (4.33\%) & 525,528 (x6.7) \\
        & sNN\hspace{5pt}   && 0.051120 \hspace{2pt} & 12.86\% & 0.047948 (6.21\%) & 249,689 (x3.2) \\
        & mNN\hspace{5pt}   && 0.050541 \hspace{2pt} & 13.85\% & 0.046899 (7.21\%) & 205,821 (x2.6) \\
       \midrule
       \multirow{4}{*}{CTRP} & dGBDT\hspace{5pt} & \multirow{4}{*}{203,650} & 0.049638 \hspace{2pt} & N/A & 0.049326 (0.63\%) & N/A \\
        & hGBDT\hspace{5pt} && 0.042638 \hspace{2pt} & 14.10\% & 0.040776 (4.37\%) & 871,621 (x4.3) \\
        & sNN\hspace{5pt}   && 0.037664 \hspace{2pt} & 24.12\% & 0.034122 (9.41\%) & 422,562 (x2.0) \\
        & mNN\hspace{5pt}   && 0.035348 \hspace{2pt} & 28.79\% & 0.030014 (15.10\%) & 321,781 (x1.6) \\
       \midrule
       \multirow{4}{*}{NCI-60} & dGBDT\hspace{5pt} & \multirow{4}{*}{675,000} & 0.055406 \hspace{2pt} & N/A & 0.055400 (0.01\%) & N/A \\
        & hGBDT\hspace{5pt} && 0.032634 \hspace{2pt} & 41.10\% & 0.031302 (4.08\%) & 12,774,917 (x18.9) \\
        & sNN\hspace{5pt}   && 0.033246 \hspace{2pt} & 40.00\% & 0.030876 (7.13\%) & 1,876,474 (x2.8) \\
        & mNN\hspace{5pt}   && 0.032020 \hspace{2pt} & 42.21\% & 0.030407 (5.04\%) & 4,138,578 (x6.1) \\
       \bottomrule
  \end{tabular}
  \label{tab:compare_scores}
\end{table}

\begin{figure}[h!]
  \centering
  \includegraphics[width=0.8\linewidth]{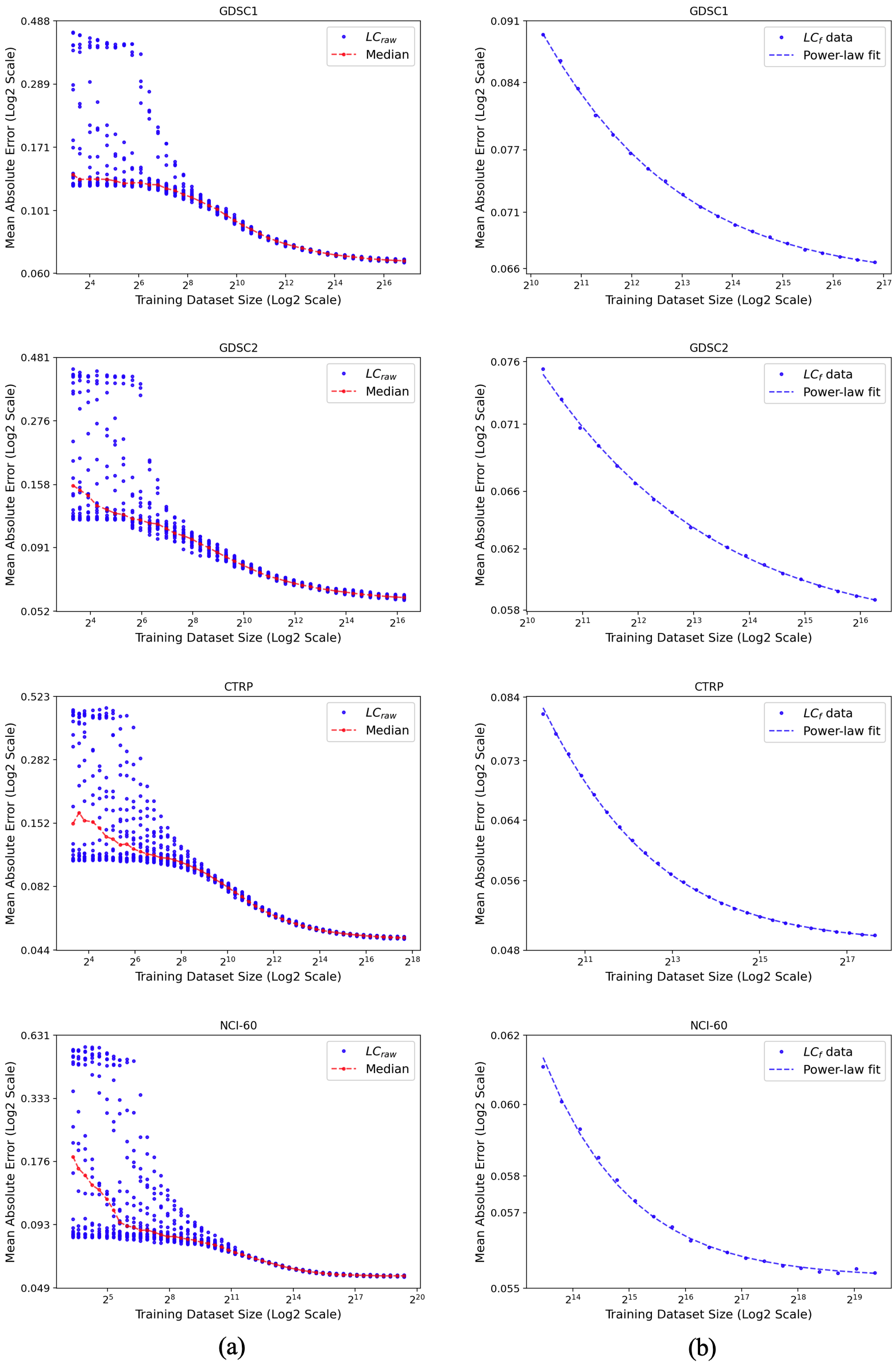}
  \caption{Learning curves generated by using dGBDT for multiple data splits of each of the datasets in Table \ref{tab:datasets}. Each data point is the mean absolute error of predictions computed on test set $E$ as a function of the training subset size $m_k$. The entire set of learning curve scores, $LC_{raw}$, is plotted in (a). A subset of scores, $LC_f$, was obtained by taking the median across the scores per size $m_k$, as shown in (b). The values in $LC_f$ were fitted to the power law model in Eq. \ref{eq:power_law}.
  }
  \label{fig:gGBDT_LC}
\end{figure}

Because of the large difference in performance between dGBDT and the other models, we analyze the learning curves of dGBDT separately in Fig. \ref{fig:gGBDT_LC}. Since LightGBM is highly parallelizable and allows faster model convergence as compared with NNs, we train 1,000 dGBDT models ($N=20$, $K=50$) with each of the four datasets, as shown in Fig. \ref{fig:gGBDT_LC}a. The scores for each subset of size $m_k$ are aggregated by computing the median (red curve).
Note that the three regions of the learning curve are apparent in each plot.

The small-data region especially stands out because of the large spread of scores on the vertical axis. The power law region follows next and can be identified as the linear region on the $log{-}log$ plot. The curve then starts to converge and progresses to the range of irreducible error. We visually identified the data points within the small-data region and excluded them from $LC_{raw}$. The remaining points, $LC_f$, were fitted with the power law in Eq. (\ref{eq:power_law}), resulting in a substantially close fit with a maximum $MAE_{fit}$ of 0.000029 and a minimum $R^2_{fit}$ of 0.99 across the four datasets, as shown in Fig. \ref{fig:gGBDT_LC}b. In each case, the curve exhibits a trajectory of convergence at higher $m_k$ values, suggesting that additional data is not expected to substantially reduce the prediction error.
These plots are the first observation demonstrating that the power law appropriately characterizes the data scaling behavior of drug response predictors.

\begin{figure}[h]
  \centering
  \includegraphics[width=0.96\linewidth]{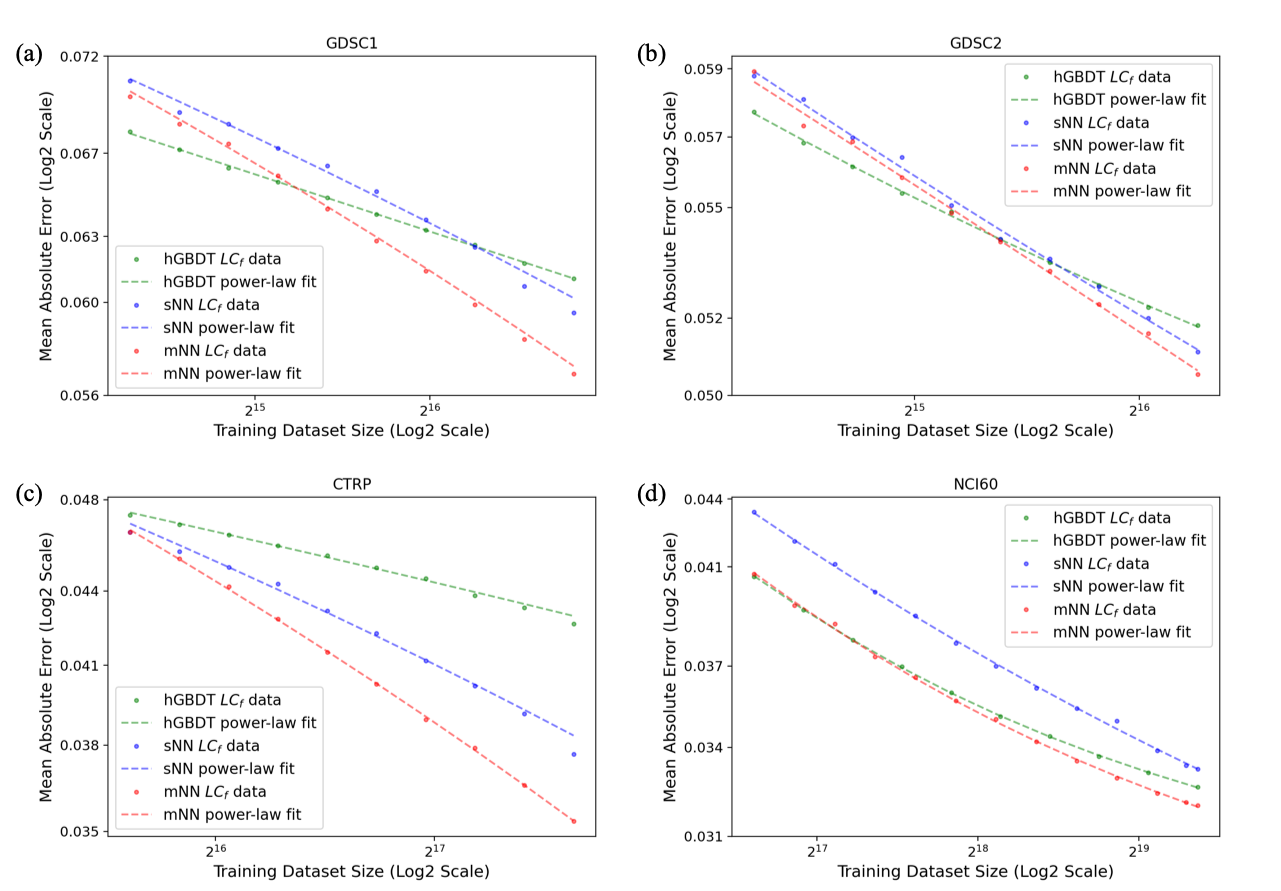}
  \caption{Comparison of learning curves of hGBDT, sNN, and mNN, for each of the four drug response datasets in Table \ref{tab:datasets}.
  }
  \label{fig:compare_lc_models}
\end{figure}

To assess the utility of learning curves as a global metric for evaluating predictors, we collected $LC_{raw}$ for the hGBDT, sNN, and mNN models for each dataset. To obtain error scores for an appropriate power-law fit, we qualitatively selected, based on empirical observations of plots in Fig. \ref{fig:gGBDT_LC}b, a range of $m_k$ that excludes the small-data region for each dataset. The selected $m_k$ range is summarized in Table \ref{tab:compare_scores} for each dataset, including the goodness-of-fit measure $MAE_{fit}$ for the power law fits. The $LC_f$ data points and the power law fits are shown in Fig. \ref{fig:compare_lc_models}.

As Fig. \ref{fig:compare_lc_models} indicates, mNN outperforms sNN across the entire range of the explored training sizes on every dataset, albeit with the similar number of trainable parameters in these NNs. This superiority of mNN can be attributed to the separate encoding of gene expressions and drug descriptors within the individual input subnetworks, enhancing the overall model learning. Moreover, mNN exhibits the lowest prediction error at the full sample size for all datasets. On both GDSC datasets, however, no single model dominates across the entire $m_k$ range: hGBDT outperforms both NNs at a lower range but performs worse than NNs as the training size increases. Another important observation is the different trajectories of the curves among the datasets. On CTRP, for example, the slope of mNN is considerably steeper than that of sNN. Thus, the mNN is expected to exhibit a higher rate of improvement on prediction score if the training size further increases. On NCI-60, however, while the NNs exhibit a similar curve for the majority of the observed range, mNN shows a sign of convergence and begins to transition from the power law region to plateau.

The power law fit can be used to address questions such as the following that allow forecasting the predictor performance beyond the available training set size.
(1) What is the expected prediction score if the training size is doubled, namely, $s(m=2|T|)$?
(2) What is the training set size required to reduce the error score by 10\%, namely, $m(s=0.9s_K)$?
These questions are addressed by plugging the appropriate values for $m_k$ or $s_k$ in Eq. (\ref{eq:power_law}) while using the power law parameter estimates, $\hat{\beta}$, for each dataset-model pair.  The rightmost two columns in Table \ref{tab:compare_scores} list the computed values addressing these two questions. The observations and results in this section directly demonstrate the benefit of using learning curves to evaluate predictors, which provide a broader perspective on the overall scaling trajectory of these models for drug response prediction, and the utility of the power-law fits for addressing important questions in prospective research.

\begin{table}[h]
\caption{Range of training set sizes in $LC_f$ used to fit the power law expression in Eq. (\ref{eq:power_law}) and the goodness-of-fit measure, $MAE_{fit}$. The $R^2_{fit}$ for all the listed experiments is higher than 0.99.}
  \centering
  \begin{tabular}{l | cc | ccc}
     \toprule
     \multirow{2}{*}{Dataset} & \multirow{2}{*}{$m_{k=1}$} & \multirow{2}{*}{$m_K=|T|$} & \multicolumn{3}{c}{$MAE_{fit}$} \\
     \cmidrule(r){4-6}
     & & & hGBDT & sNN & mNN \\
     \midrule
     GDSC1  &  20,000 & 115,863 & 0.000035 & 0.000264 & 0.000202 \\
     GDSC2  &  20,000 &  78,423 & 0.000040 & 0.000129 & 0.000123 \\
     CTRP   &  50,000 & 203,650 & 0.000103 & 0.000204 & 0.000053 \\
     NCI-60 & 100,000 & 675,000 & 0.000023 & 0.000070 & 0.000065 \\
     \bottomrule
  \end{tabular}
  \label{tab:compare_fits}
\end{table}

\section{Discussion} \label{sec:discussion}


The analysis across sixteen experiments (with four datasets and four models) demonstrates that no single model dominates in terms of prediction performance across all datasets and training sizes. This result supports the idea that the actual shape of learning curves depends on both the dataset and the model.
For example, hGBDT exhibits lower error scores for a lower range of training sizes on two datasets, as shown in Fig. \ref{fig:compare_lc_models}(a, b). Alternatively, mNN outperforms all the investigated models (dGBDT, hGBDT, sNN) across all datasets for sufficiently large training sizes. Moreover, both NNs maintain data scaling properties that are characterized by the power law region, demonstrating a promising trajectory of further improvement. These observations indicate that the power law fits can be used to project the expected error score beyond the available training size or, alternatively, calculate the sample size required to achieve a specific performance. These uses of learning curves can aid in collaboration between experimental biologists and computational scientists to shape a global vision of how predictive models can be further improved. This valuable perspective can guide the process of new data generation, either through lab experiments or synthetically, via simulations or resampling methods.

One should be cautious, however, in generalizing the data scaling characteristics when building predictors with subsets of the investigated datasets or modified architectures of sNN or mNN. In this study, we explored a general case where both cell and drug features predict the drug response. Alternatively, models that focus on a specific cancer type are usually trained by using an appropriate subset of cell lines of a much smaller sample size. Moreover, to mitigate overfitting with the reduced sample size, researchers may choose to limit the feature space by using the drug features only. These dataset and model changes may produce a different layout of learning curves and, therefore, different conclusions and downstream actions. We therefore recommend generating learning curves for every dataset-model pair that is being analyzed.

In Fig. \ref{fig:compare_lc_models}(b, d), we observe a moderate improvement at the full training set size with mNN as compared with hGBDT. Similarly, certain studies proposing novel DL methods for drug response prediction demonstrate moderate improvement as compared with classical ML \cite{Rampasek2019a, Ciriano2019Kekule, Chiu2019a}. This is in contrast to vision and text applications where DL methods represent the state of the art \cite{tan2020CVPR, howard2018ULMFiT}. The complexity in design and training of models such as mNN and hGBDT is higher than that of their respective counterparts, sNN and dGBDT. While using simpler models as a demonstration vehicle might be tempting, such models typically result in a poor baseline for objectively evaluating the prediction performance of proposed models. Similarly, generating learning curves requires significant computational resources for performing a thorough analysis across multiple data splits and training sizes, as demonstrated in this study. While often time- and resource-consuming, a rigorous comparison of novel models with strong baselines is necessary for producing a significant impact and visibility within the ML community.


Similarly to learning curve studies from other scientific domains, we demonstrate that the power law in Eq. (\ref{eq:power_law}) closely models the data scaling for the application of drug response prediction.
While other works focus primarily on a single family of ML models, we have investigated both classical ML and DL models, with a primary observation that no single model is superior to other models for all datasets and training sizes. Moreover, the extent of the three learning regions (described in Fig. \ref{fig:lrn_crv_intro}) significantly differs among the different applications. For example, Mukherjee et al. \cite{Mukherjee2003} accurately fit the power law of eight cancer-related classification tasks with DNA microarray datasets ranging between 53 and 280 samples. The prediction of drug response in cancer cell lines is presumably a more challenging task, since our models require thousands of training samples to reveal the learning regions.

While cell lines remain a primary environment for mimicking cancer, alternative biological models are being investigated as closer surrogates of human cancer. These alternatives include patient-derived xenografts (PDXs), which are cancer implants in animals, and patient-derived organoids (PDOs), which are 3-D cultures of cancer cells from patients. ML analysis with these emerging cancer environments is essential for future development of cancer treatment. At this point, however, drug response data for these biological models is scarce compared with the relatively abundant cell line screening data. The scarcity of PDXs and PDOs data imposes a challenging search for suitable methods across the entire space of learning algorithms. Therefore, learning curves can serve as a useful co-design tool for comparing predictive performance of learning algorithms and facilitate the design of future experiments in a prospective research setting.

\section{Conclusions} \label{sec:conclusion}

We demonstrate that learning curves of drug response predictors using both classical ML and DL methods follow a power law expression. The specific trajectory of the curves depends on the dataset and the learning algorithm and therefore should be obtained empirically. While hGDBT exhibits superior performance at a lower range of training sizes, the mNN outperforms all the investigated models as the training size increases. The power law can be utilized to forecast the behavior of learning curves beyond the available training size. The fitted power law curve provides a forward-looking metric for analyzing prediction performance and can serve as a co-design tool to guide experimental biologists and computational scientists in the design of future experiments in prospective research studies.

\section*{Code}
The code is available at \url{github.com/adpartin/dr-learning-curves}.

\section*{Funding}
This work has been supported in part by the Joint Design of Advanced Computing Solutions for Cancer (JDACS4C) program established by the U.S. Department of Energy (DOE) and the National Cancer Institute (NCI) of the National Institutes of Health. This work was performed under the auspices of the U.S. Department of Energy by Argonne National Laboratory under Contract DE-AC02-06-CH11357, Lawrence Livermore National Laboratory under Contract DE-AC52-07NA27344, Los Alamos National Laboratory under Contract DE-AC5206NA25396, and Oak Ridge National Laboratory under Contract DE-AC05-00OR22725. This project has also been funded in whole or in part with federal funds from the NCI, National Institutes of Health (NIH), under Contract No. HHSN261200800001E. The content of this publication does not necessarily reflect the views or policies of the Department of Health and Human Services, nor does mention of trade names, commercial products, or organizations imply endorsement by the U.S. Government.

\bibliographystyle{unsrt}  
\bibliography{arxiv_article}  

\end{document}